\documentclass[twocolumn,preprintnumbers,nofootinbib]{revtex4-2}

\usepackage{graphicx}% Include figure files
\usepackage{dcolumn}% Align table columns on decimal point
\usepackage{bm}% bold math
\usepackage{amsmath,amssymb,amsfonts}
\usepackage{latexsym}
\usepackage{physics}

\usepackage{hyperref}
\hypersetup{%
setpagesize=false,
 bookmarksnumbered=true,%
 bookmarksopen=true,%
 colorlinks=true,%
 linkcolor=blue,
 citecolor=magenta,
}

\def\nn    {\nonumber}
\DeclareUnicodeCharacter{039B}{\Lambda}

\def\nn    {\nonumber}

\begin{document}

%\preprint{APS/123-QED}

\title{\boldmath
% Goldstone preheating in $R^3$ modified $R^2$-Higgs inflation after CMB+BAO data
Echoes of $R^3$ modification and Goldstone preheating in the CMB-BAO landscape
}

\author{Tanmoy Modak}
\affiliation{Department of Physical Sciences, Indian Institute of Science Education and Research Berhampur, Berhampur 760003, India}

\begin{abstract} 
The $R^2$ and the single-field-like regime of $R^2$-Higgs inflation are disfavored by the observed high spectral index $n_s$ from the combined cosmic microwave background (CMB) and baryon acoustic oscillation (BAO) measurements at the $\sim2\sigma$ level. The addition of a dimension-six $R^3$ term in the action helps alleviate this tension. We show that the parameter space accounting for the observed high $n_s$ also induces rapid Goldstone and Higgs preheating. The preheating, especially from Goldstone modes, helps match the CMB and inflationary scales, which in turn supports the observed $n_s$.

\end{abstract}

\maketitle

%-----------------------------------------------------------------------------------------------------------------------------------
%	Introduction
%-----------------------------------------------------------------------------------------------------------------------------------

\section{Introduction}
The amplitude of the scalar power spectrum $A_s$ and the spectral index $n_s$ in the baseline $\Lambda\mathrm{CDM}$ model provide powerful tests of inflationary dynamics. Combined with the tensor-to-scalar ratio $r$, the parameters $A_s$ and $n_s$ have ruled out several inflationary models using the Planck 2018 data~\cite{Planck:2018jri}. The Starobinsky or the $R^2$ inflation~\cite{Starobinsky:1980te,Starobinsky:1983zz,Vilenkin:1985md}, in which Einstein gravity is extended by an additional $R^2$ term, is one of the best-fit models to the Planck data. However, recent combined  measurements (CMB-SPA+DESI) from the South Pole Telescope (SPT)~\cite{SPT-3G:2025bzu}, the Atacama Cosmology Telescope (ACT)~\cite{AtacamaCosmologyTelescope:2025blo}, Planck and, baryon acoustic oscillation data from Dark Energy Spectroscopic Instrument (DESI)~\cite{DESI:2025zgx}, found $n_s = 0.9728 \pm 0.0027$~\cite{SPT-3G:2025bzu}. This combined measurement is consistent with $n_s = 0.9752 \pm 0.0030$, found separately by the ACT collaboration using ACT, Planck (with lensing), and DESI-DR2 BAO data~\cite{AtacamaCosmologyTelescope:2025blo}.

The marginalised high value of $n_s$ excludes the $R^2$ model and the single-field regime of $R^2$-Higgs inflation at $>2\sigma$~\cite{SPT-3G:2025bzu,AtacamaCosmologyTelescope:2025blo}. In single-field attractor-type models, the predicted spectral index is $n_s = 1 - 2/\mathcal{N}_*$, where $\mathcal{N}_*$ is the number of $e$-folds required for the reference scale (typically set at $k_{\rm ref}=0.05,\mathrm{Mpc}^{-1}$) to exit the horizon before the end of inflation. The parameter $\mathcal{N}_*$ depends explicitly on the post-inflationary reheating history, and with Standard Model (SM) field content, $\mathcal{N}_*$ in the $R^2$ and single-field-like regime of $R^2$-Higgs inflation lies within $[50,60]$ $e$-folds. This translates to $n_s \sim [0.9600\text{--}0.9667]$ and manifests as observed CMB-BAO tension~\cite{Drees:2025ngb,Kallosh:2025ijd}. Several alternatives~\cite{Drees:2025ngb,Kallosh:2025ijd,Aoki:2025wld,Gialamas:2025kef,Aoki:2025wld,Zharov:2025evb,Liu:2025qca,Haque:2025uis,Yogesh:2025wak,Gialamas:2025ofz,Addazi:2025qra,Pallis:2025nrv,Saini:2025jlc,Wolf:2025ecy,Wang:2025dbj,Piva:2025cqi,SidikRisdianto:2025qvk,Zharov:2025zjg,Ferreira:2025lrd,Ellis:2025ieh,Odintsov:2025jky,Oikonomou:2025htz,Park:2025upd,Odintsov:2025eiv,Pozdeeva:2025wsl,Qiu:2025uot,McDonough:2025lzo,Pineda:2025ubm} have been explored to reconcile the $R^2$ and $R^2$-Higgs models with the CMB-BAO tension, with the addition of an $R^3$ term being one of the simplest~\cite{Kim:2025dyi,Addazi:2025qra,Modak:2025bjv,Park:2025upd}. In particular, it has been shown that the inclusion of an $R^3$ term can alleviate the tension in the $R^2$-Higgs model, especially when one deviates mildly from the single-field-like regime~\cite{Modak:2025bjv}.

In this letter, we show that Goldstone preheating in the $R^3$-modified $R^2$-Higgs inflation plays a quintessential role in explaining the observed high value of $n_s$. We demonstrate that Goldstone preheating (or equivalently, the production of longitudinal gauge bosons) induces parametric resonance leading to efficient production of Goldstone bosons. This results in significantly faster preheating than that of Higgs field alone. This rapid preheating improves the matching between inflationary and CMB scales and thereby helps alleviate the tension. We remark that all analyses of the $R^2$-Higgs model addressing the CMB-BAO tension thus far have adopted the unitary gauge, in which the Goldstone bosons are removed from the dynamics. We show that this gauge choice is ill-defined during every zero crossing of the Higgs background condensate after the end of inflation. By instead adopting a proper gauge choice, such as the Coulomb gauge, and employing a doubly covariant formalism for the scalar-field dynamics including all SM bosonic fields, we show that the gauge-invariant quantum energy densities of the produced Goldstone and Higgs quanta correctly reproduce the matching between the post-inflationary CMB reference scale $k_{\rm ref}$ and the inflationary scale.

\vskip0.08cm
%
%%%%%%%%%%%%%%%%%%%%%%%%%%%%%%%%%%%%%%%%%%
\section{ The action and inflationary dynamics}
%%%%%%%%%%%%%%%%%%%%%%%%%%%%%%%%%%%%%%%%%%
The action of $R^3$ modified $R^2$-Higgs inflation~\footnote{We remark that in general $f(R)$ theory additional dimension-six operators, such as $\Phi^2 R^2$ and $\Phi^6$, can also modify the inflationary predictions~\cite{Lee:2023wdm}. Moreover, additional dimension-four terms involving  $ R_{\mu\nu}R^{\mu\nu} $, $R_{\mu\nu\rho\sigma}R^{\mu\nu\rho\sigma}$ and the Gauss-Bonnet term $R^{2} - 4  R_{\mu\nu}R^{\mu\nu} + R_{\mu\nu\rho\sigma}R^{\mu\nu\rho\sigma}$ are also possible. These terms
along with their corresponding dimension-six operators may also modify inflationary predictions (See e.g. Refs.~\cite{Satoh:2007gn,Weinberg:2008hq,Guo:2009uk,Odintsov:2018zhw,Nojiri:2019dwl,Kawai:2021edk,Koh:2023zgn}).  These effects are not considered here and being studied elsewhere.}
 in Jordan frame is
\begin{align}
  S_J  =& \int d^4 x \sqrt{-g_J} \bigg[ \frac{M_{\rm P}^{2}}{2} \bigg(R  + \frac{\xi_R \ R^2}{2 M_{\rm P}^2}  + \frac{R^3}{3 M_{\rm P}^4 \xi_c } + \nn\\
  & \frac{2\xi_H}{M_{\rm P}^2} |\Phi|^2 R\bigg)
  -g_J^{\mu\nu}(\nabla_\mu\Phi)^\dagger \nabla_\nu\Phi -
  \lambda|\Phi|^4  - \nn\\
  &\dfrac{1}{4} g_J^{\mu\rho} g_J^{\nu\sigma} B_{\mu\nu}B_{\rho\sigma}- \dfrac{1}{4} g_J^{\mu\rho} g_J^{\nu\sigma} W^i_{\mu\nu}W^i_{\rho\sigma}
 \bigg],\label{eq:actionJ}
\end{align}
where $M_{\rm P}=\sqrt{1/\left(8\pi G\right)}\approx 2.4\times 10^{18}~\text{GeV}$ is the reduced Planck mass, $G$ is Newton's constant and $\sqrt{-g_J}$ is the determinant of the mostly-plus metric. The $\Phi$ is the hypercharge $+1$ Higgs field, $\nabla_\mu = D_\mu + i g' \frac{1}{2} Q_{Y} B_\mu + i g \, \bm{T} \cdot \bm{W}_\mu$ with $D_\mu$ is covariant space-time derivative. The $g'$ and $g$ are the $U(1)_Y$ and $SU(2)_L$ couplings, $Q_{Y}$ is $U(1)_Y$ hypercharge and $\bm{T}$ are the weak isospin. Here, $\xi_R$ and $\xi_c$ are the dimension-four and -six self coupling of the Ricci scalar and $\xi_H$ is the nonminimal Higgs coupling. In the baseline $R^2$-Higgs model the term associated with $\xi_c$ is zero.

A scalar degree of freedom $\phi$ emerges in the so-called Einstein frame simply by rescaling Eq.~\eqref{eq:actionJ} with $g^{\mu\nu}_J = \Theta \ g^{\mu\nu}_E$, where $\Theta=e^{\sqrt{\frac{2}{3}}\frac{\phi}{M_{\rm P}}}$. It is however easy to study the dynamics in the Einstein frame where the inflationary potential take form
\begin{align}
V_E=& \frac{1}{\Theta^2}\Biggl[\lambda |\Phi^\dagger \Phi|^2 + \frac{M_{\rm P}^4 \xi_c^2 }{48}  (\xi_R- \tilde{\zeta})^2
(\xi_R+2 \tilde{\zeta}) \Biggr], \label{eq:pot}\\
&\mbox{with}~\tilde{\zeta}=\Biggr\{\xi_R^2 + \frac{4}{\xi_c}  \bigg[\Theta -1 - \frac{2 \xi_H (\Phi^\dagger \Phi)}{M_{\rm P}^2}\bigg]\Biggl\}^{1/2}. \nn
\end{align}
The $\phi$ and, the Higgs with decomposition $\Phi=(h + i \phi_2, \phi_3+ i \phi_4)^T$, constitute a field-space manifold $\phi^I(x^\mu)\in\{ \phi, h, \phi_2, \phi_3, \phi_4\}$
with field-space metric components $G_{\phi\phi} = 1$, $G_{ h h} = e^{-\sqrt{\frac{2}{3}}\frac{\phi}{M_{\rm P}}}$, $G_{\phi_i \phi_i} = e^{-\sqrt{\frac{2}{3}}\frac{\phi}{M_{\rm P}}}$ with $i=2,3,4$. The presence of $G_{hh}$ and $G_{\phi_i \phi_i}$ is well known in the case of $R^2$-Higgs inflation and manifests itself as noncanonical kinetic terms~\cite{Langlois:2008mn,Kaiser:2012ak,Amin:2014eta,Sfakianakis:2018lzf}.
In the $\xi_c \to \infty$ limit $\tilde{\zeta}$ reduces to
\begin{equation}
\tilde{\zeta}
\simeq
\xi_R
+
\frac{2}{\xi_c \, \xi_R}
\left(
\Theta - 1
-
\frac{2 \xi_H \, \Phi^\dagger \Phi}{M_P^2}
\right).
\end{equation}
leading to to the baseline $R^2$-Higgs inflation potential as
\begin{align}
V_E(\phi^I) =& e^{-2\sqrt{\frac{2}{3}}\frac{\phi}{M_{\rm P}}} \bigg[\frac{\lambda}{4} \left(h^2+\sum^4_{i=2} \phi_i^2\right)^2
+ \frac{M_{\rm P}^4}{4 \xi_R}\Biggl\{1 -\nn\\
& \qquad e^{\sqrt{\frac{2}{3}}\frac{\phi}{M_{\rm P}}} + \frac{\xi_H}{M_{\rm P}^2} \left(h^2+\sum^4_{i=2} \phi_i^2\right)\Biggr\}^2\bigg]
\label{def:VE-final}.
\end{align}

The fields $\phi^I(x^\mu)$ are decomposed into homogeneous background part ${\varphi}^I $ and perturbation $\delta\phi^I$ as
$\phi^I(x^\mu) = \varphi^I(t) + \delta\phi^I(x^\mu)$, where only $\phi$ and $h$ acquire background field values $\varphi^I(t)=\{\phi_0(t), h_0(t)\}$
while Goldstone $\phi_2$, $\phi_3$ and $\phi_4$ are treated purely as perturbations. Here $t$ is cosmic time. The equations of motion (EoMs) for backgrounds are
\begin{align}
&\mathcal{D}_t \dot{\varphi}^I + 3 H\dot{\varphi}^I + G^{I J} V_{E,J}= 0\label{eq:bkg_inf}.
\end{align}
Here, $H=\dot{a}/a$ is the Hubble parameter and $a$ is scale factor. For any arbitrary  vector $A^I$ in field-space, the covariant time derivative is defined as
$\mathcal{D}_t A^I  = \dot{A}^I  + \Gamma^I_{\; JK} \dot{\varphi}^J A^K$ where  $\Gamma^I_{\; JK}$ are field-space
Christoffel symbol. Here $t$ is interchangeably used with number of $e$-folding before end of inflation as $\mathcal{N} \equiv \ln a(t)-\ln a_{\rm end}$.
The background energy density is $\rho_{\mathrm{inf}} = \frac{1}{2} G_{IJ} \dot{\varphi}^I \dot{\varphi}^J + V_E$
where $G_{IJ}$ and $V_E$ are evaluated at the background order. Inflation ends when $\epsilon = - \dot{H}/H^2$ equals to 1.

The perturbations $\delta\phi^I$ are gauge dependent but the Mukhanov-Sasaki variables $Q^I = \delta\phi^I + (\dot{\varphi}^I/H)\psi$
are gauge independent, where $\psi$ is scalar perturbation from the metric~\cite{Sasaki:1986hm,Mukhanov:1988jd,Mukhanov:1990me} as given as
\begin{align}
ds^2 =& -(1+2\mathcal{A}) dt^2 +2 a(t) (\partial_i \mathcal{B}) dx^i dt +\nn\\
&
a(t)^2 \big[(1-2\psi) \delta_{ij}+  2 \partial_i \partial_j \mathcal{E}\big] dx^i dx^j.\label{eq:frwmetric}
\end{align}
In the following we choose longitudinal gauge where the scalar perturbations $ \mathcal{B}$ and $\mathcal{E}$ vanish.
The EoMs for $Q^I$ are~\cite{Gong:2011uw,Kaiser:2012ak,Amin:2014eta,Sfakianakis:2018lzf,Cado:2024von,Cado:2023zbm}
\begin{align}
&\mathcal{D}_t^2 Q^{I} + 3 H \mathcal{D}_t Q^{I} -\frac{\partial^2}{a^2} Q^{I} + \mathcal{M}^{I}_{\ \ J} Q^{J}+ \mathcal{F}_{(\phi_I)}= 0,\label{eq:perturb}
\end{align}
where, $\mathcal{M}^{I}_{\ L}$ are evaluated at the background order. The $\mathcal{F}_{(\phi)}=\mathcal{F}_{(h)}=0$, while
$\mathcal{F}_{(\phi_2)}= g_Z [(1/\sqrt{6}M_{\rm P})\,\dot{\phi_0} h_0 Z_0-\dot{h}_0 Z_0+ (h_0/2)(D_\nu Z^\nu)]$,
with the replacements $g_Z\to i e/(2\sqrt{2}s_W)$ and, $Z_0\to (W^-_0-W^+_0)$ and $(iW^-_0+iW^+_0)$ for 
$\mathcal{F}_{(\phi_3)}$ and $\mathcal{F}_{(\phi_4)}$, respectively. The $s_W$ is the sine of the Weinberg angle. As only $\phi$ and $h$ acquire background values, the EoMs of $Q^{\phi_2}$, $Q^{\phi_3}$ and $Q^{\phi_4}$ decouple from $Q^{\phi}$ and $Q^{h}$. The power spectrum of the curvature perturbation $\mathcal{P}_{\mathcal{R}}$ and $n_s$ are
\begin{align}
 \mathcal{P}_{\mathcal{R}}(t;k)= \frac{k^3}{2\pi^2}\left|\frac{ H}{\dot{\sigma}} Q_\sigma\right|^2, ~~n_{s} = 1 + \frac{d\ln\mathcal{P}_{\mathcal{R}}(k)}{d\ln k},\label{eq:PRns}
\end{align}
where $\dot{\sigma} = \sqrt{G_{IJ} \dot{\varphi}^I \dot{\varphi}^J}$, $\hat{\sigma}^I = \dot{\varphi}^I/\dot\sigma$ and $Q_\sigma = \hat{\sigma}_I Q^I$.

%%%%%%%%%%%%%%%%%%%
%%%%%%%%%%%%%%%%%%%
\begin{figure*}[htbp]
\begin{center}
\includegraphics[width=.47 \textwidth]{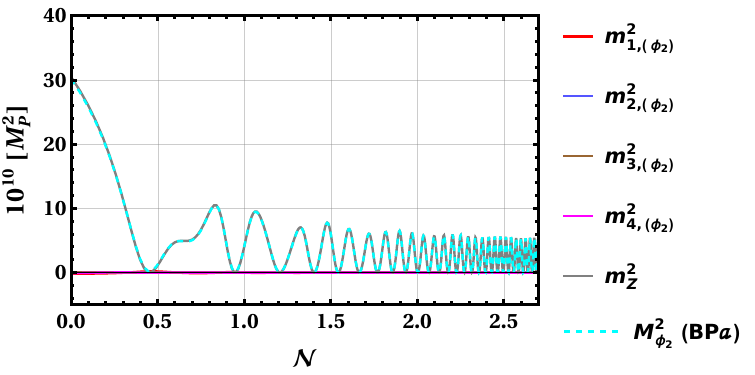}
\includegraphics[width=.49 \textwidth]{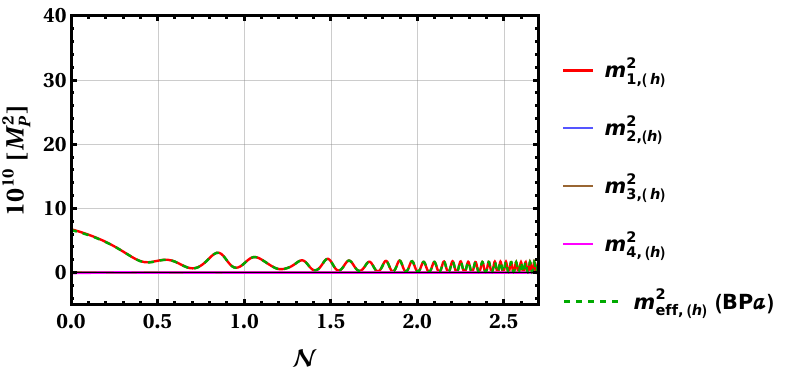}\\
\includegraphics[width=.47 \textwidth]{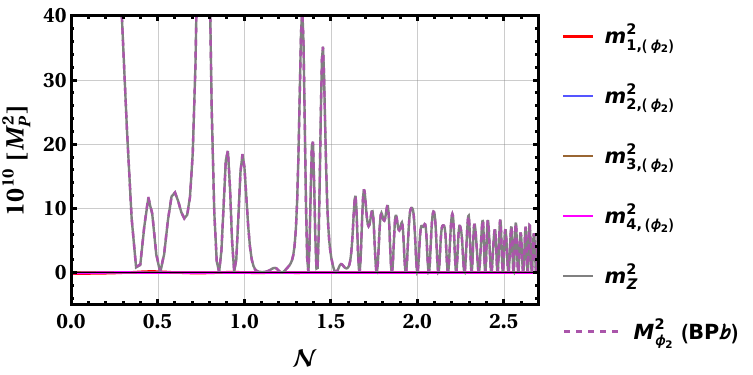}
\includegraphics[width=.49 \textwidth]{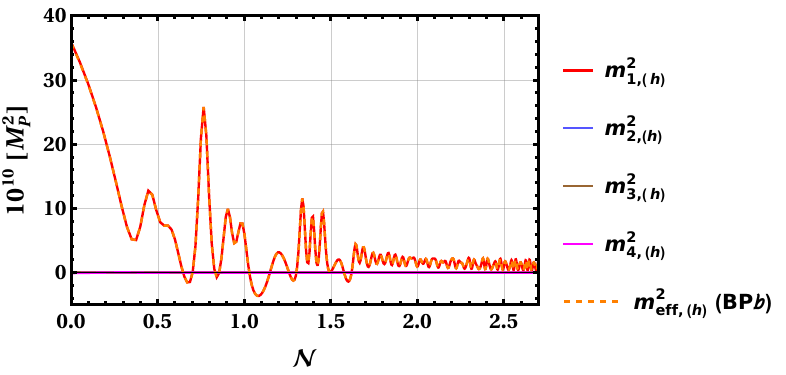}
\end{center}
\caption{The $M_{\phi_2}^2$ (left) and $m_{\mathrm{eff},(h)}^2$ (right) and, their individual components as function of $\mathcal{N}$ for the BP$a$ (upper panel) and BP$b$ (lower panel).}
\label{plot:meffplot}
\end{figure*}

We now consider a few benchmark points (BPs)~\footnote{Note that, in general, a negative value of $\xi_c$ can
reproduce the measured value of $n_s$, as discussed in Ref.~\cite{Addazi:2025qra,Park:2025upd}. A detailed study of the case with negative $\xi_c$ is currently being carried out elsewhere.}
in Table~\ref{parmeterchoices}, mimicking a
single-field-like regime with mild deviations as in Ref.~\cite{Modak:2025bjv}, to examine whether they can yield a high $n_s$.
%%%%%%%%%%%%%%%%%%%
\begin{table}[h]
\begin{tabular}{|c |c| c| c| c | c | c| c | c |c| c| c| c}
    \hline
	BP                  & $\xi_R$             &  $\xi_H$   & $\xi_c$                 &  $\phi_0(t_{\text{in}})$ [$M_{\rm P}$]  & $h_0(t_{\text{in}})$  [$M_{\rm P}$] \\
   \hline
        $a$             & $2.19\times 10^9$   &  $1.5$     & $1\times 10^{-14}$      &  5.305                                    & $8\times10^{-5}$ \\
        $b$             & $2.3\times 10^9$    &  $10$      & $8\times 10^{-15}$      &  5.35                                    & $3\times10^{-7}$ \\
	\hline
	\end{tabular}
	\caption{Two chosen BPs with $\lambda=0.01$ for both the BPs.}
	\label{parmeterchoices}
\end{table}
%%%%%%%%%%%%%%%%
The background EoMs Eq.~\eqref{eq:bkg_inf} are solved  with the initial conditions in Table~\ref{parmeterchoices}. The EoMs of the perturbations in Eq.~\eqref{eq:perturb} on the other hand is solved in momentum space with Bunch-Davies (BD) initial conditions. The solutions are then inserted the into Eq.~\eqref{eq:PRns} to obtain $\mathcal{P}_{\mathcal{R}}$ and $n_s$. We find $\mathcal{P}_{\mathcal{R}}(k_*)=2.128\times10^{-9}$ $(2.104\times10^{-9})$ and $n_{s_*}=0.9712$ $(0.9733)$ for BP$a$ (BP$b$), with $k_*=3.0498\times10^{-5}~M_{\rm P}$ $(1.3843\times10^{-4}~M_{\rm P})$ and $\mathcal{N}_*=\ln a(t_*)-\ln a_{\rm end}=-53.7$ $(-54.5)$, where $t_*$ corresponds to cosmic time when the reference mode $k_*$ exits horizon. These values lie within $\sim1\sigma$ of the CMB-SPA+DESI constraints $\log(10^{10}A_s)=3.0574\pm0.0094$ and $n_s=0.9728\pm0.0027$ \cite{SPT-3G:2025bzu}. The single-field estimate of $r\simeq16\epsilon(\mathcal{N}_*)$ yields $r\simeq3.59\times10^{-3}$ $(3.45\times10^{-3})$, consistent with the 95\% CL BICEP/Keck bound \cite{BICEP:2021xfz}. Matching $k_*$ to the reference scale $k_{\rm ref}$ is deferred for later section since it depends on the post-inflationary reheating history. We note that one can define an effective single-field potential $W_{\rm eff}(\phi)$ by solving $\frac{\partial V_E}{\partial h_0}=0$ for $h_0$ and substituting back into $V_E$ to compute the background and perturbation EoMs. However, Ref.~\cite{Modak:2025bjv} shows that the resulting $n_s$ can differ significantly from that obtained using the full potential $V_E$. We also verified that isocurvature perturbations in this approach are about three orders of magnitude smaller for the reference mode and remain suppressed during inflation. We refer the reader to Ref.~\cite{Modak:2025bjv} for these details.
\vskip0.08cm

%
%%%%%%%%%%%%%%%%%%%%%%%%%%%%%%%%%%%%%%%%%%
\section{Preheating}
%%%%%%%%%%%%%%%%%%%%%%%%%%%%%%%%%%%%%%%%%
In the early part of reheating (i.e. preheating), the rapidly oscillating condensate can produce particles nonperturbatively and modify the thermal history, thereby affect in the matching. We show shortly that the BPs in Table~\ref{parmeterchoices} can induce successful preheating. To study preheating, a gauge choice is required. While the unitary gauge is commonly used in the $R^2$-Higgs model, it becomes ill-defined at each zero crossing of $h_0$ after inflation, as understood easily from Eq.~\eqref{eq:perturb} and from the explicit expression of $\mathcal{F}_{(\phi_2)}$~\cite{Sfakianakis:2018lzf,Cado:2024von}. Same is true for $\phi_3$ and $\phi_4$. We therefore adopt the Coulomb gauge, $\partial_i Z^i=0$ and $\partial_i W^{\pm i}=0$, which is well defined for all background values~\cite{Sfakianakis:2018lzf,Cado:2024von}. In this gauge one may treat either the Goldstone modes or the longitudinal gauge bosons as dynamical; in the following we keep the Goldstone modes dynamical and focus on $\phi_2$. The dynamics of $\phi_3$ and $\phi_4$ are analogous.

The perturbation is first rescaled $X^{\phi_2} \equiv a \ Q^{\phi_2}$ and then quantized in momentum space by
\begin{align}
\hat{\widetilde{X}}^{\phi_2} = s_k(\tau) e^{\phi_2}(\tau) \hat{a}(\vb{k}) + s^*_k(\tau) e^{\phi_2}(\tau) \hat{a}^\dagger(-\vb{k}), \label{eq:phi2quant}
\end{align}
where $\hat{\widetilde{X}}^{\phi_2}$ is the Fourier transformed $X^{\phi_2}$. The $\hat{a}^\dagger$, $\hat{a}$ are creation and annihilation operators, $s_k$ is the mode function, $e^{\phi_2}$ is the vielbein with $e^{\phi_2}e^{\phi_2}=G^{\phi_2\phi_2}$, and $\tau$ is conformal time ($\partial_0\to\partial_\tau/a$). The decoupled mode equation is given as~\cite{Sfakianakis:2018lzf,Cado:2024von}
\begin{align}
& s_k'' + \frac{ 2 m_Z^2 \Upsilon}{\mathcal{K}_Z} s_k' + \bigg[ k^2  + a^2 M_{\phi_2}^2+ \frac{ 2 m_Z^2 \Upsilon^2}{\mathcal{K}_Z}\bigg] s_k  = 0,\label{eq:goldmode}
\end{align}
where $m_Z^2=  (g_Z^2/4)  e^{-\sqrt{\frac{2}{3}}\frac{\phi_0}{m_{\rm P}}} h_0^2$, $M_{\phi_2}^2=m_{\mathrm{eff},(\phi_2)}^2+ m_Z^2$ and $(\prime)$ denotes
the conformal time derivative. The $\mathcal{K}_Z$  and $\Upsilon$  are given as
\begin{align}
&\mathcal{K}_Z =\frac{k^2}{a^2}+ m_Z^2,~~\Upsilon (\tau) = \frac{\phi_0'}{\sqrt{6}M_{\rm P}} -  \frac{a'}{a} -   \frac{h_0'}{h_0}.
\end{align}
The effective mass $m_{\mathrm{eff},(I)}^2$ is given as
\begin{align}
m_{\mathrm{eff},(I)}^2(\tau)= \mathcal{M}^I_{~~I} - \frac{1}{6}R_E G^I_{~I},\label{eq:meffsq}
\end{align}
where $R_E= -(2 H^2+\dot{H})$ and
\begin{align}
&\mathcal{M}^{I}_{\ L} = G^{IJ} (\mathcal{D}_L\mathcal{D}_J V_E)- \mathcal{R}^I_{\ JKL} \dot{\varphi}^J \dot{\varphi}^K\nn\\
&\qquad  \qquad \qquad \qquad \qquad - \frac{1}{M_{\rm P}^2 a^3} \mathcal{D}_t \left(\frac{a^3}{H}\dot{\varphi}^I \dot{\varphi}_L\right),\label{eq:massterm}
\end{align}
One can identify different components of $m_{\mathrm{eff},(I)}^2$ as
\begin{subequations} \begin{eqnarray}
m_{1,(I)}^2 &=& G^{(I)J} (\mathcal{D}_{(I)}\mathcal{D}_J V_E),  \label{def:effective-masses-decomposition-1} \\
m_{2,(I)}^2 &=& - \mathcal{R}^{(I)}_{\ \ JK(I)} \dot{\varphi}^J \dot{\varphi}^K, \label{def:effective-masses-decomposition-2} \\
m_{3,(I)}^2 &=& - \frac{1}{M_{\rm P}^2 a^3} \mathcal{D}_t \left(\frac{a^3}{H}\dot{\varphi}^{(I)} \dot{\varphi}_{(I)}\right), \label{def:effective-masses-decomposition-3} \\
m_{4,(I)}^2&=&-\frac{R_E}{6},  \label{def:effective-masses-decomposition-4}
\end{eqnarray}  \label{def:effective-masses-decomposition}   \end{subequations}
with $(I)$ indices are not summed such that
\begin{align}
m_{\mathrm{eff},(I)}^2= \sum_k m_{k,(I)}^2.
\end{align}
We have utilized  here the Coulomb gauge condition $\partial_i Z^i=0$ and rescaled $Z_0 \to Z_0/a$. The mode equation for the Higgs is found by setting terms with $\Upsilon$ to zero and the replacement $M_{\phi_2}^2 \to m_{\mathrm{eff},(h)}^2$. The corresponding $M_{\phi_2}^2$ and $m_{\mathrm{eff},(h)}^2$ along with their respective contributions are plotted in Fig.~\ref{plot:meffplot}. The dominant contribution to $M_{\phi_2}^2$ comes from $m_Z^2$, while $m_{1,(I)}^2$ dominates $m_{\mathrm{eff},(h)}^2$. Since both effective masses are evaluated at the background level and the Goldstone bosons are treated as perturbations, they receive no significant contribution from $V_E$. For the considered BPs, $M_{\phi_2}^2$ is much larger than $m_{\mathrm{eff},(h)}^2$.

The vacuum-subtracted quantum energy density for the Goldstone mode $\phi_2$ is~\cite{Cado:2024von}
\begin{align}
\rho^q_{(\phi_2)} = \frac{1}{a^4}\int\left( \frac{d^3k}{(2\pi)^3} \rho_k^{(\phi_2)} - \frac{k^3}{4\pi^2 \Delta_{(\phi_2)} } dk \right), \label{eq:quantumphi}
\end{align}  
where  
\begin{align}
\Delta_{(\phi_2)} = \exp{\int_{-\infty}^{\tau}\frac{ 2 m_Z^2 \Upsilon}{\mathcal{K}_Z}\,d\tau'},
\end{align}  
and  
\begin{align}
\rho_k^{(\phi_2)}\; &= \;   \frac{1}{2} \Biggl\{ \left(1-\frac{m_Z^2}{\mathcal{K}_Z} \right)\left|s_k^\prime\right|^2
 +  \Biggl[k^2+a^2 m_{\mathrm{eff},(\phi_2)}^2 \nn\\
&- \frac{m_Z^2}{\mathcal{K}_Z} \;\Upsilon^2\Biggl] \left|s_k\right|^2
- \frac{m_Z^2}{\mathcal{K}_Z} \;\Upsilon \; (s'_k s^\ast_k +s^{\ast\prime}_k s_k )
\Biggr\}.\label{eq:golphi2}
\end{align} 
We compute $\rho^q_{(\phi_2)}$ by solving Eq.~\eqref{eq:goldmode} with BD initial conditions $s_k =e^{-i k \tau}/(\sqrt{2k\Delta_{(\phi_2)}})$ and $s_k^\prime$. in cosmic time via initializing all modes deep inside the horizon~\cite{Sfakianakis:2018lzf,Cado:2024von}. The $\Delta_{(\phi_2)}$ factor in the denominator appears due to the presence of the second friction term in Eq.~\eqref{eq:goldmode}. Note that we solved Eq.~\eqref{eq:goldmode} in cosmic time for computational convenience. The second term in the integrand of Eq.~\eqref{eq:quantumphi} is associated with the BD energy density for the corresponding mode. A mode is excited if $\rho_k^{(\phi_2)}$ exceeds the corresponding BD energy density.

%%%%%%%%%%%%%%%%%%%
%%%%%%%%%%%%%%%%%%%
\begin{figure}[h]
\begin{center}
\includegraphics[width=.4\textwidth]{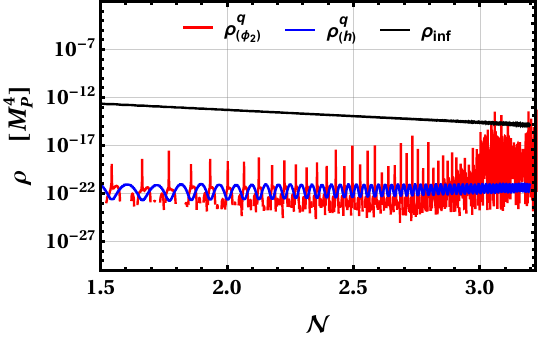}
\includegraphics[width=.4 \textwidth]{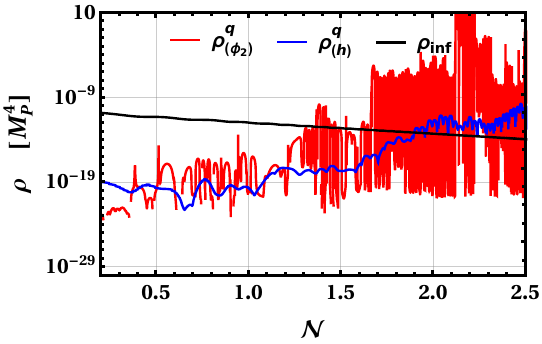}
\end{center}
\caption{The $\rho^q_{(h)}$ (blue) and $\rho^q_{(\phi_2)}$ (red) along with
$\rho_{\rm inf}$ (black) for BP$a$ (upper panel) and BP$b$ (lower panel) respectively.}
\label{plot:energy}
\end{figure}

In Fig.~\ref{plot:energy}, we show the quantum energy densities of the Higgs and Goldstone modes, $\rho^q_{(h)}$ (blue) and $\rho^q_{(\phi_2)}$ (red), along with $\rho_{\rm inf}$ (black). For both BPs, $\phi_2$ triggers preheating first; at $\mathcal{N}_{\rm pre}\simeq 3.2$ and $1.4$ respectively. Higgs preheating is incomplete for BP$a$, but completes for BP$b$ $\sim2$ $e$-folds after inflation. Preheating is deemed complete if a perturbation energy density is equal to $\rho_{\rm inf}$~\cite{Cado:2024von}. Note that the Goldstone preheating is faster due to larger $M_{\phi_2}^2$ and the $\Upsilon$-$\mathcal{K}_Z$ term in contrasts to only $m_{\mathrm{eff},(h)}^2$ in case of Higgs. The Goldstone preheating is faster for BP$b$ primarily due to larger $\xi_H$. Likewise, Higgs preheating is possible for BP$b$ due to its larger $m_{\mathrm{eff},(h)}^2$. Here we take $\xi_H \sim 10$ as a representative value. However, larger values $\xi_H \gtrsim 10$ are also viable and, with successful preheating, may also assist in matching the scale.

In the absence of the $R^3$ term, the model reduces to the baseline $R^2$-Higgs potential of Eq.~\eqref{def:VE-final}, as discussed above. While preheating may still
occur for the ballpark values of $\xi_R$ and $\xi_H$ for the BPs~\cite{Cado:2024von} but, one cannot account
the observed large $n_s$ in the absence $R^3$ term~\cite{Modak:2025bjv}.
Preheating from $\phi_3$ and $\phi_4$ is nearly identical to $\phi_2$, differing only by $g_Z \to i e/(2 \sqrt{2} s_W)$. We find preheating from $\phi$ is weak, while transverse $Z$ and $W^\pm$
modes take longer. This finding is similar to Ref.~\cite{Cado:2024von}. For $\xi_H<1$, Goldstone as well as gauge boson preheating is subdued. In such scenario, nonvanishing $\xi_c$ can still explain high $n_s$, but additional fields are needed for preheating. Otherwise, thermalization proceeds via perturbative reheating~\cite{Modak:2025bjv,Ferreira:2025lrd}.

Assuming that  the thermalization is immediately completed after preheating, we estimate preheating temperature $T_{\rm pre}$ via
\begin{align}
\rho_{\mathrm{inf}}\bigr|_{\mathcal{N}=\mathcal{N}_{\rm pre}}\equiv\rho_{\rm pre}=\frac{g_{\rm pre} \pi^2}{30} \; T_{\rm pre}^4,\label{eq:prehettemp}
\end{align}
where $g_{\rm{pre}}=106.75$ is the number of relativistic degrees of freedom at the completion of preheating.
We find that $T_{\rm pre} \approx 2.2 \times 10^{14}~(7.8 \times 10^{14})$ GeV for BP$a$ (BP$b$).
This emphasizes the role of $\xi_c$ and $\xi_H$ in accounting the observed $n_s$ and also thermalization process via preheating.

%%%%%%%%%%%%%%%%%%%
\vskip0.08cm
%
%%%%%%%%%%%%%%%%%%%%%%%%%%%%%%%%%%%%%%%%%%%%%%%%%%%%%%%%%%%%%%%%%
\section{Inflationary observables and scale matching}
%%%%%%%%%%%%%%%%%%%%%%%%%%%%%%%%%%%%%%%%%%%%%%%%%%%%%%%%%%%%%%%%
We proceed now matching of $k_*$ (in $M_{\rm P}$) to the CMB reference scale $k_{\rm ref}/a_0 = 0.05~\rm{Mpc}^{-1}$, where $a_0$ is scale factor today. The CMB reference scale is defined as $k_{\rm ref} = k_* = a_* H_*$, giving~\cite{He:2020ivk}
\begin{align}
\mathcal{N}_* = -\ln\Biggl[\frac{H_*}{k_{\rm ref}/a_0} \frac{T_0}{T_{\rm pre}} \frac{g_0^{1/3}}{g_{\rm pre}^{1/3}}\Biggr] + \mathcal{N}_{\rm pre}, \label{eq:refsca}
\end{align}  
where $T_0=2.7~\rm{K}$, $g_0=43/11$, and $a_*, H_*$ are evaluated at $\mathcal{N}_*$.
The two sides of Eq.~\eqref{eq:refsca} must coincide for exact matching of the CMB and inflationary scales. Using the values of $\mathcal{N}_{\rm pre}$, $T_{\rm pre}$, $H_*$, and $a_*$, the RHS vs LHS yields $-54.8$ ($-55.3$) vs $-53.7$ ($-54.5$) for BP$a$ (BP$b$). This is within $\sim1$ $e$-fold between two sides of Eq.~\eqref{eq:refsca} for the BPs.  Therefore, the matching is very close, though not exact. Of course, exact matching could be achieved here by adjusting $\xi_R$, $\xi_c$, and $\xi_H$, but is deferred due to neglected condensate decays, rescattering, and perturbation decay effects, which require a nonlinear analysis. These issues will be addressed elsewhere.

%%%%%%%%%%%%%%%%%%%
\vskip0.08cm
%
%%%%%%%%%%%%%%%%%%%%%%%%%%%%%%%%%%%%%%%%%%
\section{Discussion and Summary}
%%%%%%%%%%%%%%%%%%%%%%%%%%%%%%%%%%%%%%%%%%
We showed that $R^3$ can make the single-field-like regime of the $R^2$-Higgs model compatible with observations. The parameter space that accounts for the observed high $n_s$ also induces strong Goldstone preheating, which in turn helps match the reference CMB scale to inflationary dynamics. Using two representative parameter sets, we find that in the $R^2$-like regime, with nonminimal coupling $\xi_H \gtrsim 1.5$, a $\xi_c \sim 10^{-14}$ is sufficient to trigger preheating. While a nonvanishing $\xi_c$ can still account for the high $n_s$, Goldstone and Higgs preheating require $\xi_H \gtrsim 1.5$. We primarily focused on the $R^2$-like regime; however, in the Higgs-like regime with a larger $\xi_H$, preheating is faster and matching becomes easier. It should also be noted that, unlike pure Higgs inflation, here the unitarity cut-off is restored up to the Planck scale, see e.g. Ref.~\cite{Ema:2017rqn} and, the produced Goldstone bosons do not violate the unitarity. This novel connection between the CMB-BAO tension, higher-order curvature effects, and preheating in $R^2$-Higgs inflation has not been addressed in the literature. We also note that our study leaves room for improvement. Primary uncertainties arise from unaccounted condensate and perturbation decays, which may require going beyond the linear-order approximation and including non-equilibrium effects.

Interpreting the CMB-BAO tension as a precursor to new physics (NP) requires caution, as a better understanding of the correlations between the datasets within the $\Lambda$CDM model is needed. The BAO data do not directly affect $n_s$, and the tension could arise from unaccounted systematics~\cite{Ferreira:2025lrd}. However, intriguingly, the combined CMB measurements from ACT, SPT, and Planck yield $n_s = 0.9684 \pm 0.0030$, slightly higher than Planck 2018’s $n_s = 0.9657 \pm 0.0040$, though the difference is small. These results may hint at NP beyond $\Lambda$CDM, but independent CMB and BAO measurements from missions like the Simons Observatory (with $\sigma(n_s) \sim 0.002$), DES, and Euclid are needed. In this regard, 21cm intensity mapping by the Square Kilometre Array may provide an independent test~\cite{Modak:2022gol}. If the tension becomes significant, it may indicate the presence of an $R^3$ term and thus provide a promising probe of quantum gravity.

% %%%%%%%%%%%%%%%%%%%
 \vskip0.08cm
% %
\noindent{\bf Acknowledgments} \
We thank Yann Cado and Evangelos I. Sfakianakis for useful discussion.

%%%%%%%%%%%%%%%%%%%
\vskip0.08cm
%
% %%%%%%%%%%%%%%%%%%%%%%%%%%%%%%%%%%%%%%%%%%
% \section{Appendix}
% %%%%%%%%%%%%%%%%%%%%%%%%%%%%%%%%%%%%%%%%%%

%

%%%%%%%%%%%%%%%%%%%%%%%%%%%%%%%%%%%%%%%%%%%%%%%%%%%%

\end{document}